
\input phyzzx
\newcount\comcount
\def\com#1#2{\advance\comcount by1{\bf #2}\footnote{*\number\comcount*}{#1}}

\def\RREF#1#2{\gdef#1{\REF#1{#2}#1}}
\def\vev{{\nonfrenchspacing v.e.v.}}
\def\jnl#1&#2(#3){\begingroup\let\jnl=\dummyj@urnal\sl #1\bf#2\rm
(\afterassignment\j@ur\count255=#3)\endgroup}
\def\PRL{ {\sl Phys. Rev. Lett.} }
\def\PR { {\sl Phys. Rev.} }
\def\NP { {\sl Nucl. Phys.} }
\def\PL { {\sl Phys. Lett.} }

\RREF\sakarov{A.D. Sakharov, JETP Lett. {\bf 6} (1967) 24}
\RREF\andiI{A.G. Cohen, D.B. Kaplan and A.E Nelson,\NP{\bf B349} (1991) 727}
\RREF\fuku{M. Fukugita and T. Yanagida, \PR {\bf D42} (1990) 1285;
 J. Harvey and M.Turner, \PR{\bf D42} (1990) 3344;
 B. Campbell, S. Davidson, J.Ellis and K. Olive, \PL{\bf B256} (1992) 457}
\RREF\manton{F.Klinkhamer and N. Manton, \PR {\bf D30} (1984) 2212}
\RREF\turok{N. Turok and J. Zadrozny, \PRL {\bf 65} (1990) 2331}
\RREF\kaiser{N.G. Deshpande, J.F. Gunion, B. Kayser and F. Olness,
 \PR{\bf D44} (1991) 837}
\RREF\kolb{E. Kolb and M. Turner, {\it The Early Universe}
 (Addison--Wesley, New York, 1990)}
\RREF\shapo{V. Kuzmin, V. Rubakov and M. Shaposhnikov, \PL{\bf 155B} (1985) 36}
\RREF\andiII{A.G. Cohen, D.B. Kaplan and A.E. Nelson,
  \NP {\bf B373} (1992) 453}
\RREF\dine {M. Dine, P. Huet, R. Singleton and L. Susskind,
  \PL {\bf B257} (1991) 351}
\RREF\shapoII{S. Khlebnikov and M. Shaposhnikov, \NP {\bf B308} (1988) 885}
\RREF\linde{M. Dine, R.G. Leigh, P. Huet, A. Linde and D. Linde,
  \PR {\bf D46} (1992) 550}
\RREF\moha{R.N. Mohapatra and X. Zhang, UMDHEP preprint
  (92--230), May 1992}
\RREF\lep{{\it Review of Particles properties}, \PR {\bf D45},
  Part 2 (June 1992)}
\RREF\andiIII{A.G. Cohen and A.E. Nelson, UCSD preprint,
  PTH 92--32, BU--HEP--92--20, August 1992}
\RREF\shapoIII{A.M. Kazarian and M.E. Shaposhnikov,
  \PL {\bf 276} (1992) 131}
\RREF\mohaII{R.N. Mohapatra in {\it``CP violation"}, C. Jarlskog ed.
  (World Scientific, Singapore, 1989) 384}
\RREF\ecker{G. Ecker and W. Grimus, \NP {\bf 258} (1985) 328}
\RREF\chang{D. Chang, \NP {\bf 214} (1983) 435}
\RREF\branco{G.C. Branco and L. Lavoura, \PL {\bf 165B} (1985) 327}
\RREF\piet{M. Pietroni, DFPD preprint,  DFPD/92/TH/36, (1992)}
\RREF\andiIV{A.G. Cohen, D.B. Kaplan and A.E. Nelson,\PL {\bf 294} (1992) 57}
\RREF\anderson{G.W. Anderson and L.J. Hall, \PR {\bf D45} (1992) 2685}
\RREF\stan{S. Myint, \PL {\bf B287} (1992) 325}
\RREF\jmfI{G.C. Branco, J.-M. Fr\`ere and J.-M. G\'erard,
  \NP {\bf B221} (1983) 317}
\RREF\jmfII{J.-M. Fr\`ere, J. Galand, A. Le Yaouanc, L. Oliver, O. P\`ene and
  J.-C. Raynal,  \PR {\bf D46} (1992) 337}
\RREF\campbell{B. A. Campbell, S. Davidson, J. Ellis and K. A. Olive,
  CERN--TH--6646/92, Alberta Thy--30--92, UMN--THH--112/92}

\nopubblock
\titlepage \line{\hfill\vbox{
  \hbox{CERN--TH 6747/92}
  \hbox{ULB--TH--07/92}
  \hbox{UAB-FT-298/92}
  \hbox{hep-ph/9301228}}}
\vskip0.2cm plus 0.2fil
\title{Generation of the Baryon~Asymmetry~of~the~Universe within the
  Left--Right Symmetric Model}
\vskip0.5cm plus 0.7fil
\centerline{
J.-M. Fr\`ere\foot{Ma\^\i{}tre de recherches FNRS}$^a$,
L. Houart\foot{Aspirant FNRS.}$^b$,
J.M. Moreno$^c$,
J. Orloff\foot{e-mail: orloff@cernvm.cern.ch}$^a$ and
M. Tytgat\foot{Aspirant FNRS, e-mail: mtytgat@ulb.ac.be}$^b$
}
\vskip 0.5cm plus 0.1fil
\centerline{\vbox{
\tenit\halign{#&#\hfil\cr
a &Theory Division, CERN, CH-1211 Geneva 23, Switzerland\cr
b &Service de Physique Th\'eorique, CP 225, Universit\'e
Libre de Bruxelles,\cr
  & Bld. du Triomphe, B-1050 Bruxelles, Belgium\cr
c &Dept. de Fisica Teorica, Univ. Autonoma de Barcelona,\cr
  &E--08193 Bellaterra, Spain\cr}}}
\vskip 0.5cm plus 0.7fil

\centerline{Abstract}
Fermions scattering off first-order phase transition bubbles, in the
framework of $SU(2)_L\otimes SU(2)_R\otimes U(1)$ models, may generate
the Baryon Asymmetry of the Universe (BAU), either at the
$LR$-symmetry-breaking scale, or at the weak scale. In the latter
case, the baryon asymmetry of the Universe is related to CP violation
in the $K_0$--$\bar K_0$ system.

\vskip 0.5cm plus 0.7fil
\line{\vbox{\hbox{CERN--TH 6747/92}
  \hbox{ULB--TH--07/92}
  \hbox{UAB-FT-298/92}
  \hbox{December 1992}} \hfil}
\endpage

\chapter{Introduction}

Generation of the BAU at a relatively low scale was a natural response
to growing fears that sphaleron-like configurations at the weak scale
would destroy any pre-established baryon asymmetry. In fact two
responses are possible and both find a natural realization in $LR$
models.

In the first case, one assumes that the existing or generated baryon
number is protected by some symmetry immune to the weak forces (a
typical exemple is conservation of $(B-L)$ as defined in terms of the
usual fermions).  This is studied in section 2 below, where a
non-vanishing $(B-L)$ is generated at the $R$ phase transition, i.e.
when large masses are induced for bosons associated with the $SU(2)_R$
group.

Alternatively, the baryon number can be generated at the usual weak
scale, while making sure that $B$-violating interactions quickly cease
to be in equilibrium, so that the newly-born BAU cannot be washed away
\refmark{\sakarov}. Several mechanisms have been proposed to generate
the baryon asymmetry at this rather low scale\refmark{\shapo, \andiI,
\andiII, \turok, \dine}. All of them involve extensions of the
Standard Model. This may seem strange, since even the minimum standard
model already possesses all the required qualitative ingredients (C
and CP violation at the standard Lagrangian level, $(B+L)$ violation
through sphaleron-like solutions and, depending upon the scalar mass,
departure from equilibrium due to a first-order phase transition). The
main reason of this failure rests in the smallness of the invariant
effective CP-violation parameter $\delta_{CP} =
O(10^{-20})$\refmark{\shapo}, where the heavy suppression results not
only from the bare CP-violating phase of the Kobayashi--Maskawa
matrix, but also from the products of mixing angles and mass
differences needed to reflect the non-degeneracy of the 3-generation
mixing structure. As $\delta_{CP}$ is expected to enter as a factor in
the calculation of the baryon asymmetry, this falls short by many
orders of magnitude. Present data thus strongly suggest and to some
extent legitimize at least some extension of the standard model.  Such
extensions, usually to the scalar sector, may seem {\it ad hoc}, and
require the introduction of new sources of CP violation, unrelated to
low-energy phenomenology. As we shall see in section 3, $LR$ models
offer a natural framework for such extensions, with a scalar structure
similar to models already suggested. The gauge bosons associated with
$SU(2)_R$ serve as intermediaries, as they amplify the effect of the
phases, making them detectable in the usual K-system $\Delta S = 2$
amplitudes \refmark{\jmfI}.

\chapter{Baryon Number Generation at the Right Scale}

We consider in this section the possibility of creating the BAU at the
$R$-symmetry breaking scale. Various mechanisms have been imagined to
produce the asymmetry using $(B-L)$-violating interactions
\refmark{\andiI}, leading to a $B-L \not= 0$ Universe.

This in turn constrains any left-over $(B-L)$-violating interactions
at lower temperatures to be out of equilibrium, to prevent them from
erasing the BAU \refmark{\fuku}. Neutrino Majorana masses are a
typical example.

In the $LR$ symmetric model, all the ingredients to generate the
baryon asymmetry are at hand. The spontaneous breakdown of the
$SU(2)_L \otimes SU(2)_R \otimes U(1)_{B-L}$ gauge symmetry to
$SU(2)_L \otimes U(1)_Y$ breaks C and $(B-L)$ (through Majorana masses
for the right-handed neutrinos). Some CP violation in the leptonic
sector is easily included as this phenomenology is rather
unconstrained: several $R$-scalar triplets with non-removable phases
between their vacuum expectation values, or two triplets and one
pseudoscalar singlet easily foot the bill. This extension is
sufficient to generate a non-vanishing $(L)$, and has little effect at
low energy as the quarks do not couple to the scalar triplets.

The model also needs some $(B+L)$-violating processes. Such processes
induced by sphaleron-like configurations at equilibrium will convert a
fraction of $L \not= 0$ to $B \not= 0$. Two qualitatively different
configurations are possible:
\item{-} the usual sphaleron configuration associated with $SU(2)_L$ or rather
its instanton-like continuation above the electroweak scale, creates or
destroys left-handed fermions;
\item{-} on the other hand the topological argument of Manton and
Klinkhamer\refmark{\manton} may be extended to the breaking of a $SU(2)_R$
gauge symmetry.

To verify this last statement, let us consider the simple model of
$SU(2)_R$ gauge bosons coupled to a triplet complex scalar field
($\Delta_R$). We thus both neglect the couplings (both of the triplet
and of the gauge bosons) to the other scalars (bi-doublets) ---which
anyway do not develop a \vev{} at the right breaking scale--- and the
mixing with the $U(1)_{B-L}$ gauge field (just as $\theta_W=0$ was
assumed in ref.[\manton]). The scalar potential then reads
$$V(\Delta_R)=
-{\mu^2 \over 2}\tr\Delta^{\dagger}\Delta
+{\lambda_1\over 4} \left(\tr\Delta^{\dagger}\Delta\right)^2
+{\lambda_2\over4} \left(\tr\Delta\Delta\right)
  \left(\tr\Delta^{\dagger}\Delta^{\dagger}\right)		\eqn\potright
$$
where
$$\Delta = \left(
  \matrix{ \delta^+\over\sqrt2 &\delta^{++}\cr
  \delta^0 &-{\delta^+\over \sqrt 2}\cr}
\right)
={\vec\tau} \cdot{\vec r}.			\eqno\eq
$$
$\vec r= \vec r_1 + i \vec r_2 $ is a 3-dimensional complex vector and
$\tau_a$ are the Pauli matrices. Expressed in term of $\vec r$,
\potright becomes
$$V(\vec r_1, \vec r_2)=
-\mu^2 ({\vec r_1}^2+{\vec r_2}^2)
+\lambda_1 ({\vec r_1}^2 + {\vec r_2}^2)^2
+\lambda_2 \left(({\vec r_1}^2-{\vec r_2}^2)^2
  + 4 (\vec r_1\cdot \vec r_2 )^2 \right)		\eqno\eq
$$

The potential is minimized by the following solutions:
$${\vec r_1}^2={\vec r_2}^2={\mu^2\over4\lambda_1}\doteq{v_R^2\over2}
\hbox{ and }\vec r_1 \cdot \vec r_2 = 0. \eqno\eq
$$

Note that the quartic term in $\lambda_2 > 0$ is necessary to break
completely the $SU(2)_R$ symmetry by giving a mass to the otherwise
pseudo-Goldstone doubly-charged scalar particles.

The topological argument of Manton\refmark{\manton} now applies. The
existence of sphalerons ---{\it i.e.} classical unstable solutions---
is related to the possibility of constructing non-contractible loops
in configuration space, beginning and ending in the vacuum. Taking the
configuration of maximal energy on every loop of a given homotopy
class and then selecting the infimum of this ensemble, yields (modulo
the validity of Morse theory on non-compact spaces) such a sphaleron.

For a given configuration on the loop to have a finite energy, the
scalar field must stay in the vacuum at spatial infinity. This defines
a mapping of the boundary of the 3-dimensional space, $S^2$
(parametrized by $\theta,\, \phi$), onto the space of the vacuum
configurations of the scalar field, which is $SO(3)$ since the
symmetry is completely broken. What is called a {\it closed loop} is
simply a continuous family of such mappings, parametrized by $\mu \in
[0,\pi]$, and shrinking to single points of the $SO(3)$ vacuum-space
for $\mu=0,\,\pi$. The homotopy classes of such loops are then
isomorphic to those of the maps
$$g(\mu,\theta,\phi): S^3 \rightarrow SO(3).\eqno\eq
$$
where $S^3$ is spanned by $\mu ,\theta, \phi$.

Since $\pi_3(SO(3)) = Z$, the necessary topological condition for the
existence of spha\-le\-rons is fullfilled. On top of this topological
argument, much hard work would still be needed to find an explicit
solution. Unfortunately indeed, the spherical symmetry of the energy
functional, used in the complex doublet case, is lost here (it was
linked to the custodial $SU(2)$ symmetry in the limit where
$\sin\theta_W \rightarrow 0$ with vanishing Yukawa couplings). This
complicates matters considerably, as the full set of coupled
non-linear partial derivative equationsmust now be tackled; we will
not pursue this study further in the present paper. Nevertheless, with
the topological conditions satisfied, we may consider that at high
temperature (well above $v_R$), both left- and right-baryon numbers
are violated, while between the $R$ scale and the electroweak breaking
scale, only ``left" configurations are active.

The last ingredient required to satisfy Sakharov's \refmark{\sakarov}
conditions is a departure from equilibrium. We consider the generation
of $B$ through the Charge Transport Mechanism of Cohen et
al.\refmark{\andiI}, i.e.  the reflection of right-handed neutrinos on
walls of expanding bubbles of $SU(2)_L \otimes U(1)_Y$
``vacuum"\foot{Another starting point was used in ref.  [\moha],
namely the out-of-equilibrium decay of right-handed Majorana
neutrinos.}. When bubble walls are thin, this mechanism is very
efficient as it exploits a large region in front of the wall for
biasing baryon number. In contrast, the mechanism known as spontaneous
baryogenesis, is only efficient in the walls, which thus have to be
thick\refmark{\turok}. We restrict this study to a strongly
first-order phase transition, and hence relatively thin walls.  This
assumption does not upset any phenomenology as the scalar sector at
the $R$ scale is anyway barely constrained.

At the $R$ scale first-order transition, a wall separates two distinct
regions : one symmetric under $SU(2)_R \otimes SU(2)_L \otimes
U(1)_{B-L}$ (the symmetric region) and the other, symmetric only under
$SU(2)_L \otimes U(1)_Y$ (the broken one). Right-handed neutrinos
incident from the symmetric region will interact strongly with the
wall and may be either reflected or transmitted. If $N_R$ stands for
the field associated with right-handed neutrinos, it also describes
the left-handed anti-neutrinos (their CP-conjugates). Those are also
present in the thermal bath, and similarly interact with the wall.
When necessary, we will use the notation $(N_R)^c$ to distinguish
them. (In this we follow the tradition, although $N_R^{CP}$ would seem
more appropriate).  With ${\cal R}_{N_R\rightarrow (N_R)^c}$ defined
as the probability of reflection of an $R$-neutrino into
anti-$R$-neutrino, we get, under the C, P and T discrete symmetries
\foot{We find it easier to picture these symmetries for fermions
incident on a spherical bubble rather than on a plane wall.}
$$\eqalign{
\hbox{P}: \ \ {\cal R}_{N_R\rightarrow (N_R)^c}& \rightarrow {\cal
R}_{N_L\rightarrow (N_L)^c} \cr
\hbox{C}: \ \ {\cal R}_{N_L\rightarrow (N_L)^c}&
\rightarrow {\cal R}_{(N_R)^c\rightarrow N_R} \cr
\hbox{T}: \ \ {\cal R}_{(N_R)^c
\rightarrow N_R}& \rightarrow {\cal R}_{N_R\rightarrow (N_R)^c}
\cr}\eqno\eq
$$
Since only right-handed fields are involved, CPT brings no
information, as it merely relates ${\cal R}_{N_R\rightarrow (N_R)^c}$
to itself. If CP is not conserved, ${\cal R}_{N_R\rightarrow (N_R)^c}$
may differ from ${\cal R}_{(N_R)^c\rightarrow N_R}$. As in [\andiI] we
introduce CP violation in the reflection on the wall through a
space-dependent unremovable phase in the potential. The Dirac
equations for $N_R$ then involves a complex, space-dependent mass
$m(z)$:
$$\eqalign{ i \gamma^{\mu} \partial_{\mu} N_R &=m(z) (N_R)^c. }\eqno\eq
$$

A detailed calculation can be found in [\andiI] in a slightly different
version, and we don't repeat it here. The flux of leptons in the symmetric
region is related to the calculated reflection probabilities as
$$\eqalign{
f_L = {2 \over \gamma} \int\limits_m^\infty dk_L \int\limits_0^\infty
dk_T
&\left ( f^s(k_L,k_T) - f^b(k_L,k_T)\right)\times\cr
&\times\left({\cal
R}(k_L)_{N_R\rightarrow
(N_R)^c} - {\cal R}(k_L)_{(N_R)^c \rightarrow N_R}\right)\cr}
\eqno\eq
$$
where $f^s$ and $f^b$ measure respectively the flux from the symmetric
and from the broken regions, obtained by boosting the thermal
equilibrium distributions from the rest frame of the plasma to the
rest frame of the moving wall.

Owing to rapid $(B+L)$-violating interactions in the symmetric region,
the generated $L$-number partially converts into a $B$ excess. If
thermal equilibrium is satisfied in the symmetric region, this yields
[\andiI]
$$(B + L)|_{eq} \ \approx x \, \left(B - L\right)|_{eq}\eqno\eq
$$
with $x = O(1)$. Actually, thermal equilibrium is not perfect and one has to
solve the Boltzman equation for the system [\andiI]. The maximum predicted
baryon-to-entropy ratio in the symmetric region is again\foot{We have
extrapolated this  number from the weak scale\refmark{\andiI} to the $LR$
symmetry-breaking scale.  Without substantial entropy production, it is
essentially scale-independent.}
$${n_B\over s}\sim {f_L\over s} \lsim 10^{-6}.\eqno\eq
$$
This is a peak value, and the exponent is very sensitive to the parameters
involved (wall speed and thickness, fermion mass, critical temperature,...),
and typical values quoted in ref.[\andiI,\andiII] range down to $10^{-10}$,
while the CP angle comes in as a mere factor.

After completion of the phase transition, $(B - L)$ is no longer
conserved. The right-handed neutrinos have acquired a Majorana mass
through coupling to the \vev{} of the right scalar triplet that
carries a $(B - L)$-charge. Even at scales much below the $R$
transition, some of those interactions feed through to the remaining
light sector, for instance by the see-saw induced light Majorana
masses of ordinary (left-handed) neutrinos. Moreover, $(B+L)_L$
violating processes are still active, and these conjugated effects
endanger the previously created asymmetry.

To simplify the discussion, we will consider an abrupt transition
between two qualitatively different equilibrium regimes. The first one
was considered above in the ``symmetric region": all interactions are
supposed to be at least approximately in equilibrium. This yields,
through the charge transport mechanism, a non-zero baryon and lepton
number excess for all the flavours of quarks and leptons. The second
regime occurs after the $R$ phase transition: there we assume the
heavy particles to be out-of-equilibrium, hence eliminating the
$R$-gauge bosons, the right-handed Majorana neutrinos and the
$R$-scalar-triplet field from the thermal bath. We are then
essentially in the regime of the Standard Model above the electroweak
breaking scale. The critical difference is the occurrence of
$L$-number-violating processes through the virtual exchange of
Majorana neutrinos. The effective cross-sections for these processes
is
$$\sigma_{\Delta L_i=2} \approx {m_{\nu_i}^2 \over {2 \pi v_R^4} .}\eqno\eq
$$
As the density of relativistic species is $n_i = T^3/\pi^2$, the rate
$\Gamma_{\Delta L=2} = \left< \sigma n \right>$ is
$$\Gamma_{\Delta L_i =2} \approx {1\over \pi^3}{T^3 m_{\nu_i}^2 \over
v_R^4}. \eqno\eq $$
In a $LR$ model with scalar triplets, the light neutrino mass $m_\nu$ occurs
through a see-saw mechanism \refmark{\kaiser}. Barring fine tuning, this
typically leads to:
$$ m_{\nu} \approx m_{Dirac}^2 / M_{R}\eqno\eq
$$
where $m_{Dirac} \approx m_{charged\ lepton}$.

Setting $m_{\nu_e} = m_e^2 / M_R$, $m_{\nu_{\mu}} = m_{\mu}^2 / M_R$
and $m_{\nu_{\tau}} = m_{\tau}^2 / M_R$, we check which of the
lepton-number-violating processes fall out of equilibrium at $T_c \sim
M_R \sim 10$ TeV, a phenomenologically relevant scale. Using
$$m_e = 0.5 \hbox{ MeV}, \  m_{\mu} = 105 \hbox{ MeV}, \  m_{\tau} = 1800
\hbox{ MeV},\eqno\eq
$$
we get the comparison with experimental numbers:
$$\eqalign{
m_{\nu_e} &\approx 10^{-10} \hbox{ GeV} \ll 17\hbox{ eV} \cr
m_{\nu_{\mu}} &\approx 10^{-6} \hbox{ GeV} \ll 0.27 \hbox{ MeV} \cr
m_{\nu_{\tau}} &\approx 10^{-4} \hbox{GeV} \ll 35 \hbox{ MeV} \cr}\eqno\eq
$$
Comparing the rates for $\Delta L_i = 2, \, i = e,\mu,\tau$, with the
expansion rate of the Universe at $T \sim T_c \sim M_R$
\refmark{\kolb}:
$$\Gamma_{\Delta L_i = 2} \leq H = 1.67 g_*^{1/2} T^2 / m_{pl}\eqno\eq
$$
where $g_* \sim 100$, we find that for $m_{\nu_i} \leq 10^{-6}$ GeV
the process is out of thermal equilibrium. This is certainly not the
case for $L_{\tau}$-changing processes but might be the case for
$L_e$- and even for $L_{\mu}$-changing reactions. An approximate
conservation in one or two of the leptonic flavours thus appears
possible, provided lepton mixing is kept under control (see also
ref.[\campbell]). One can rest on low-energy phenomenology to argue
that those mixings are much smaller than in the quark sector.

We thus consider a thermal-equilibrium situation with conserved $L_e$
and $L_{\mu}$ and non-conserved $L_{\tau}$. Writing all the processes
in equilibrium (as are the $(B+L)_L$-violating processes) generates a
set of relations between chemical potentials (see for example Harvey
and Turner in ref.  [\fuku]) which leads to
$$(B+L)|_{eq} = - { 288 N + 30 m + 141 \over 288 N +78 m -147}
(B-L)|_{eq}\eqno\eq
$$
where $N$ and $m$ count the number of left scalar triplets and scalar
bi-doublets still participating in the thermal bath (we expect $N
\rightarrow 0$ if explicit $LR$ symmetry is imposed).  This number
must be understood as the maximum value the asymmetry may reach, since
we have neglected non-equilibrium effects between the two regimes,
which can only diminish $B$.

In conclusion, it is possible to generate the baryon number at the $R$
scale through the breaking of $(B-L)$ and both left- and
right-$(B+L)$-violating processes. This scale does not need to be very
high ---$O$(10 TeV)--- as conservation of at least one lepton number
suffices to protect the baryon number.  The predictive power of this
$R$-scale mechanism is unfortunately rather poor, as it requires the
introduction of new CP violation in the leptonic sector, without
significant low-energy implications. We now turn to the more
challenging possibility of creating $B$ at the lower electroweak scale
within a $LR$-symmetric structure.

\chapter{ Baryon Number Generation at the Weak Scale }

The main excitement about this model is that the CP-violating
parameters needed in the scalar sector are transferred by $R$ gauge
bosons to the $K$ system and thus become accessible to experiment. As
we shall see, this model also meets another challenge faced by baryon
number generators in their panic fear of thermodynamical equilibrium
situations. It is clear indeed that, unless some protection mechanism
exists ({\it e.g.} non-vanishing $(B-L)$, as discussed in the previous
section) such thermal equilibrium will tend to wipe out any existing
or generated baryon number. The way to prevent a freshly generated
baryon number from such fate is thus to require that the phase
transition occurs briskly, putting the system out of equilibrium. In
particular, the order parameter $v(T)$ must move directly to a large
value. The baryon-number-violating processes are then quickly turned
off at the onset of the phase transition. In the standard model, this
translates, as we recall below, into a small value for the mass of the
symmetry-breaking scalar. This low-mass scalar can however be avoided
if trilinear couplings between the relevant fields are present in the
tree Lagrangian, as is the case in the $LR$ model with spontaneous CP
violation.

Consider the Boltzman factor governing the rate of baryon-number
violation $\Gamma_B$:
$$\Gamma_B = 3 \kappa \alpha^4 T^4 \exp(-E_{sph}/T),\eqno\eq
$$
where $\kappa$ incorporates uncertainties in the prefactor ($\kappa=O(1)$)
and $E_{sph}$ is the energy of the saddle configuration known as the
sphaleron.  This energy is related to the $T$-dependent order
parameter $v(T)$
$$E_{sph} = \alpha_W(T) \, v(T) \,
B\left(\lambda(T)\over\alpha_W(T)\right).\eqno\eq $$

The numerical value of $B(\lambda/\alpha)$ varies from $1.5$ to $2.7$
when the ratio of the quartic scalar coupling to the gauge parameter
$\lambda / \alpha$ varies from $0$ to $\infty$. On the other hand, the
value of $v(T_c)$ at the first-order phase transition is
$E/\lambda(T_c)$ where $E$ is the trilinear coupling coefficient in
the effective potential. In the standard model this coupling is absent
at tree level, and $E = O(g^3)$. A big jump in $v$ at the phase
transition thus imposes a small $\lambda$, whence, as
$m_{higgs}\approx \lambda\, v(T=0)$, a small scalar mass.

Imposing $\Gamma_B \leq H$, where $H$ is the expansion rate of the
Universe at the time of the phase transition leads to the constraint
[\shapoII],[\linde]:
$$mass_{Higgs} \lsim 35\hbox{ GeV},\eqno\eq
$$
which lies ``uncomfortably close" to the experimental bound at LEP1
\refmark{\lep},
$$mass_{Higgs} \gsim 48\hbox{ GeV}.\eqno\eq
$$

Considerations of this kind have been applied to extensions of the
Standard Model such as the two-Higgs model [\shapoII] and the minimal
supersymmetric model [\stan]. Typically, this stills constrains
(sometimes very strongly [\stan]) the mass of the lightest scalar. A
very interesting model from this point of view is the non-minimal
supersymmetric model considered in ref.[\piet]. The constraint on the
Higgs mass is considerably relaxed by the appearance of a ``tree
level" effective cubic coupling. As seen above, the large size of the
trilinear coupling removes the requirement of a tenuous quartic
coupling. We will argue that the $LR$-symmetric model shares the same
nice feature.

However, one thing struck us in most of these constructions, namely
the closeness of the proposed models to the scalar structure
traditionnally used in the $SU(2)_L \otimes SU(2)_R \otimes U(1)$ $LR$
model. Imbedding the proposed schemes in a fully-fledged $LR$
structure would at the same time offer some justification for the
doubling of the scalars, and relate baryon-number violation to more
mundane CP parameters\foot{We wish to thank A.  Cohen for numerous
discussions on this point}. What came as a surprise is that the
further requirement of spontaneously broken CP also provided the
necessary ingredients for a tree-level driven first-order phase
transition.  Even independently of the extra nicety of this
first-order phase transition, the $LR$ model, as remarked by Mohapatra
\refmark{\moha}, offers to our knowledge the only link between our
existence as ``baryon beings" and the (minute) value of CP violation.

Unfortunately, as we shall see below, this link is not so strong as to
predict {\it e.g.} the sign of $\epsilon ' / \epsilon$ from the
dominance of matter over antimatter. This is due to the fact that even
the most straightforward model of spontaneous CP violation in $LR$
still allows for (discrete) number of variants, between which the data
and the knowledge of the strong matrix elements are currently too
unprecise to choose. Nevertheless, it is only a matter of time before
this is clarified, and the uniqueness of the link persists.

This situation is in contrast with existing models where the CP
violation related to the BAU is disconnected from the $K$-system, and
at most affect the electric dipole moment of quarks\refmark{\andiIII}
and leptons\refmark{\shapoIII}.

\section{A sketch of the $LR$ model with spontaneous CP violation}

The model we consider is CP-conserving \refmark{\mohaII, \chang,
\ecker} before symmetry breaking of the electroweak scale. Spontaneous
CP violation appears trough a non-zero physical phase in the \vev{} of
the bi-doublet fields:
$$\eqalign{\left < \phi \right > &= {e^{i {\alpha \over 2}} \over \sqrt{
2}}
\left(\matrix{\kappa & 0 \cr 0 & \kappa'\cr}\right) = {1 \over \sqrt 2}
\left(\matrix{v & 0 \cr 0 & w\cr}\right)\cr \left < \tilde \phi \right > &=
{1\over \sqrt{ 2}} \left(\matrix{w^* & 0 \cr 0 & v^*\cr}\right).\cr}\eqno\eq
$$
This, through Yukawa couplings to the quarks:
$$\eqalign{M^{(u)} &= {1 \over \sqrt 2} ( v \Gamma + w^* \Delta ) = U
D^{(u)}
U^T\cr M^{(d)} &= {1 \over \sqrt 2} ( w \Gamma + v^* \Delta ) = V D^{(d)}
V^T,\cr}\eqn\GammaDelta
$$
leads to CP violation in the $K_0$--$\bar K_0$ system. One may note
the particular form of the diagonalization of the mass matrices that
results from imposing explicit CP conservation before symmetry
breaking. In this case, the $\Gamma$ and $\Delta$ matrices can be
chosen to be real and symmetric.

In a particular basis for the quarks fields, the KM matrices for the L and R
charged currents read
$$K_L = U^\dagger V, \ \ \ \ \ \ K_R = K_L^*\eqno\eq
$$

which are thus not independent matrices. In this convention all the phases in
$K_L$ are observable: their number is ${n(n+1) \over 2}$ for $n$ generations,
hence six for the case of interest, all of them related to $\alpha$ (of course,
for two generations all the phases and for three generations all the phases but
one, can be rotated away from $K_L$ into $K_R$ but this does not make them
unobservable). This is the zero-temperature situation. To look at the behaviour
of $\alpha$ at non-zero temperature, more information is needed on the scalar
potential, which is unfortunately not very constrained in the $LR$ model. It
contains many unknown couplings between the bi-doublet and the two triplets.
One can still show \refmark{\branco} that, without fine-tuning, this potential
is unable to generate a non-zero value for the phase $\alpha$ after SSB. This
is easy to cure through the introduction of a singlet neutral pseudo-scalar
[\branco], which couples to the bi-doublet in the following way
$$V(\Phi,\eta) = V (\Phi) + V_{\eta}(\Phi, \eta),\eqno\eq
$$
$$V_{\eta}=V(\eta)+iC_1\eta \left({\rm det}\Phi-{\rm det}\Phi^*\right)+
C_2\eta^2 tr\left(\Phi^\dagger \Phi\right).\eqno\eq
$$
Once expressed in polar coordinates, the trilinear term in the potential
depends upon $\sin \alpha$ and will compete with terms even under $\alpha
\rightarrow -\alpha$, leading to $\alpha\not=0$ for a large range of
parameters.

\section{Phase Transition and Baryogenesis}

We will not attempt here to pursue the study of the above potential at
finite temperature.  Nevertheless reliable hints can be obtained from
simpler models.  In particular we now argue, in analogy with [\piet],
that a strongly first-order phase transition may take place.

The argument is based on the existence of a trilinear term (required
for natural spontaneous CP violation) in the classical potential. The
Higgs field, i.e. the field that develops a non-zero v.e.v., is a
temperature-dependent combination of $\phi_1^0$, $\phi_2^0$ and
$\eta$, with
$$\eqalign{ \left<\phi_1\right>&=\left<H\right>\cos\beta_1\cos\beta_2\cr
\left<\phi_2
\right> &= \left< H \right> \cos \beta_1 \sin \beta_2 \cr
\left<\eta\right> &= \left< H \right> \sin
\beta_1,\cr}\eqno\eq
$$

where $\beta_1$ and $\beta_2$ are temperature-dependent mixing angles.
These angles will remain sizeable provided the masses and interactions
of the involved scalars are not too different. This requires a
relatively light $\eta$, associated with the $L$ scale rather than the
$R$ scale, as is more usually assumed.

The orthogonal combinations correspond to massive scalars with zero \vev{}
Once substituted in the potential $V_\eta$, \? yields a term trilinear in
$H$. We have thus good indications that a strongly first-order transition will
arise, without seriously constraining the masses of the light scalars. We refer
the reader to ref. [\piet] for an explicit exemple in a more tractable model.

A strongly first-order phase transition must proceed through the
nucleation of bubbles with thin walls \refmark{\anderson}. The
spontaneous baryogenesis mechanism of [\moha] uses $\left < \dot
\alpha \right>$ as a source biasing baryon-number-violating processes.
In the present case of a thin wall, we turn to the more efficient
charge transport mechanism of Cohen, Kaplan and Nelson
\refmark{\andiII}.

They consider the diffusion of a fermion (the $top$ quark, as its
Yukawa coupling is the largest) by the expanding bubble of true
vacuum. An incident $top$ (massless in the false vacuum) of right
chirality ($t_R$) will be reflected as a $top$ of left chirality
($t_L$) or transmitted as a massive $top$ of right helicity\foot{One
may also expect to see the $top$ transform into a $bottom$, which
interacts much less with the wall. However it is easy to find a gauge
(so-called unitary gauge) in which there is no stationary gauge field
configuration. Any $top$ to $bottom$ transition must then be due to
virtual effects in the presence of the wall and thus be reduced by
some effective $G_W$, a much smaller effect than the tree-level effect
of diffusion considered in [\andiII].}. Under C,P and T, $t_R$
transforms as:
$$\eqalign{
\hbox{P}&: \ \ {\cal R}_{t_R\rightarrow t_L} \rightarrow {\cal
  R}_{t_L\rightarrow t_R},\cr
\hbox{C}&: \ \ {\cal R}_{t_L\rightarrow t_R}
  \rightarrow{\cal R}_{(t_R)^c\rightarrow (t_L)^c},\cr
\hbox{T}&: \ \ {\cal R}_{(t_R)^c\rightarrow
  (t_L)^c} \rightarrow {\cal R}_{(t_L)^c\rightarrow (t_R)^c}. \cr}\eqno\eq
$$

The CPT theorem imposes ${\cal R}_{t_R\rightarrow t_L} = {\cal
R}_{(t_L)^c\rightarrow (t_R)^c}$. Only if CP is conserved do we get
${\cal R}_{t_R\rightarrow t_L}\buildrel \it CP \over = {\cal
R}_{(t_R)^c\rightarrow (t_L)^c}\buildrel \it CPT \over = {\cal
R}_{t_L\rightarrow t_R}$ . If the interaction with the wall violates
CP, those two probabilities may differ, as shown in [\andiII]. In the
rest frame of the wall, an observer in the symmetric region sees a
flux of particles with a non-zero axial baryon number:
$$\eqalign{
f_A = {2 \over \gamma} \int\limits_m^\infty dk_L \int\limits_0^\infty
  dk_T
 &\left ( f^s(k_L,k_T) - f^b(k_L,k_T)\right)\times\cr
&\times\left({\cal R}(k_L)_{t_L\rightarrow
 t_R} - {\cal R}(k_L)_{t_R \rightarrow t_L}\right)\cr}\eqno\eq
$$
where $f^s$ and $f^b$ refer respectively to the boosted flux from the
symmetric and from the symmetry-broken region. The flux $f_A$ is non
zero if CP is violated and if the wall is moving through the thermal
background.  This flux carries no net baryonic charge, but is easily
shown to carry a hypercharge\foot{Actually $f_Y = {1\over 4} f_A$. It
has been critized that Debye screening of the gauged hypercharge would
prevent this current to penetrate far in the symmetric region, thus
curbing the mechanism. One should however keep in mind that this
current, while carrying some hypercharge, is not identical to the
latter, but merely refers to the hypercharge in the top sector.
Screening of hypercharge can then occur with the help of all quarks
equally without seriously hindering the mechanism, as shown in
[\andiIV]. For brevity we keep referring to an excess $Y$.}.

As hypercharge is conserved in the symmetric region (as well as $Q$,
$B-L$, ...)  but $B+L$ is not\foot{The rate of baryon-number violation
in the symmetric region has been estimated on dimensional ground to be
of order $\Gamma_B \approx \kappa \alpha_W^4 T^4$ where $\kappa =
O(1)$.}, equilibrium processes will lead to a non-zero baryon density
excess in front of the wall.  After completion of the phase
transition, this excess of $B$ is transferred to the broken symmetry
Universe.

The authors of [\andiII] obtain in this way a baryon to entropy ratio
$$B \buildrel def\over = \,\left.n_B \over s\right|_{pred.} \approx 10^{- 8}
\Delta
\alpha \eqno\eq
$$
where $\Delta \alpha$ is the jump of $\alpha$ from the symmetric to
the broken region\foot{The above result is of course subject to many
uncertainties related, on one side to rough estimates concerning
baryon-number-violating processes, and on the other side to variations
depending on the wall velocity and width themselves. }. Below we will
take the approximation $\Delta\alpha = \alpha (T=0)$. We argued indeed
that the phase transition may be strongly first order, which means
that the jump in the order parameters must be close to their $T=0$
value.

The above estimate of $B$ agrees with the value obtained from
nucleosynthesis\refmark{\kolb}
$$4 \times 10^{-11} \leq n_B / s \leq 1.4 \times 10^{-10}.\eqno\eq
$$
even for values of $\Delta \alpha$ as small as $0.01$ to $0.001$.
Larger values of $\Delta \alpha$ could of course be accommodated if
some later dilution of $B$ occurs (for instance, if the $B$-violating
processes are not fully out of equilibrium in the broken phase).

\section{Connection to Low-Energy CP violation}

It is interesting to compare the $T=0$ value of $\alpha$ obtained
above with the low-energy CP violation. While the comparison should be
made to the full model\refmark{\jmfII}, it is informative to first
consider the simple case of two quark generations, where the
dependence on the various parmeters is considerably easier to exhibit.
The $K_L$ matrix is parametrized in the following form:
$$K_L = e^{i\gamma} \left(\matrix{e^{-i \delta_2} \cos \theta_c
& e^{-i \delta_1}\sin\theta_c\cr
-e^{i \delta_1} \sin \theta_c & e^{i \delta_2}
\cos \theta_c\cr}\right).					\eqno\eq
$$
The phases appearing in this matrix can be related to $\alpha$ and to the
measured values of quarks masses and mixings. An exact treatment [\jmfII]
largely corrobrates Chang's linear development [\chang] which yields
$$\eqalign{
\delta_1 &\approx {r \sin\alpha \over 1- r^2 \cos^2 \alpha}\left
 (-{3\over 2} {m_c\over m_s} \cos^2 \theta_c - {1 \over 2}{m_c\over m_d}
 \sin^2\theta_c \right )\cr
&\cr
\delta_2 &\approx {r \sin\alpha \over 1- r^2 \cos^2
 \alpha}\left (-{1\over 2} {m_c\over m_s} \cos^2 \theta_c + {1 \over 2}
 {m_c\over m_d}\sin^2 \theta_c \right)\cr}\eqno\eq
$$
where $r =|\kappa'/\kappa|$.
With $m_s \approx 200$ MeV, $m_c \approx 1.4$ GeV, $m_d \approx 8.9$
MeV, and small $\alpha$, this reduces to
by baryogenesis considerations one gets:
$$ |(\delta_1 - \delta_2)| \approx 10\, |r \alpha|. \eqno\eq
$$ Those phases can be related in the minimal $LR$ model to the experimentally
well--known value of $\epsilon$:
$$\eqalign{
\epsilon_{LR}&\approx e^{i\pi/4}\,0.36\,
  \sin(\delta_1-\delta_2)\,({1.4\hbox{ TeV}\over M_R})^2,\cr
\epsilon_{exp} &= e^{i\pi/4}\, 2.26\, \times 10^{-3}.\cr}\eqno\eq
$$

Some bounds on $r$ arise from the inversion procedure, which
re-expresses $\Delta$ and $\Gamma$ (cf. eq. \GammaDelta) in terms of
the measured quark masses and mixings (for details see [\jmfII]). One
typically gets (if we choose arbitrarily $\kappa'$ to be the smaller
of these parameters)
$$r = O \left(m_b\over m_t\right).\eqno\eq
$$
It is to be noted that this limit on $r$, imposed by the existence of
the third generation, is a matter of mathematical consitency and must
be taken into account even if dominant contributions to the process
considered come from the first two generations.

It is then clear that the values of $\alpha$ required to account for
$\epsilon$, even in a simple 2-generation scheme, are if anything more than
sufficient to account for the observed baryon number of the Universe, thus
allowing for some dilution with respect to the mechanism studied.

This is obviously not the complete story, and even the simplest $LR$
model with spontaneous CP violation allows much more freedom in the
relation between these parameters. In particular, a discrete set of
models is associated with the choice of the relative signs of the
masses in ref. [\ecker]: while in the Standard Model we can always
redefine $q_R \rightarrow - q_R$ without observable consequences, this
is not possible in the $LR$ model. Furthermore, the range of allowed
values is considerably larger in the full 3-flavour study ---this
being in part due to the lack of precise measurements for mixing
angles and the unknown mass of the $top$.

A detailed study of the various observable CP-violating parameters as
a function of $\alpha$ ($\epsilon$, $\epsilon'$ and the neutron
electric dipole moment) is presented in [\jmfII], to which we refer
the reader. From that study it can be seen that the values of $\alpha$
needed for high-temperature CP violation are perfectly compatible with
those observables, but that critical tests (e.g.  the agreement of the
sign of $\epsilon' \over \epsilon$) will require more precise
measurements of the Kobayashi--Maskawa mixing angles.

\chapter{Conclusions}

The minimal $LR$ model considered in this paper has many attractive
features for the phenomenology of CP violation both at low energy and
at high temperature.  To our knowledge it offers the first
link\refmark{\moha} between previously disconnected sectors: the
generation of the baryon number of the Universe and the value of
$\epsilon$ from the $K^0$--$\bar K^0$ system, even if too many
uncertainties (both theoretical and experimental) temporarily prevent
us from fully exploiting the real predictive power of the model.

\ack
We wish to thank the Orsay group for a collaboration on the $LR$ model
to which we refer extensively in the present paper; Andy Cohen, Belen
Gavela, Olivier P\`ene and Philippe Spindel for uncountable and
on-going discussions; JMF also wishes to thank the theory group of
BNL, where part of this work was completed.

\refout
\end